\documentclass[%
 reprint,
 amsmath,amssymb,
 prstab,
floatfix,
]{revtex4-2}

\usepackage{graphicx}
\usepackage{dcolumn}
\usepackage{bm}
\usepackage{cleveref}

\begin{document}

\preprint{APS/123-QED}

\title{Slice energy spread measurement in the low energy photoinjector}%

\author{Houjun Qian}
\email{houjun.qian@desy.de}
\affiliation{Deutsches Elektronen-Synchrotron, 15738 Zeuthen, Germany}%

\author{Mikhail Krasilnikov}
\email{mikhail.krasilnikov@desy.de}
\affiliation{Deutsches Elektronen-Synchrotron, 15738 Zeuthen, Germany}%

\author{Anusorn Lueangaramwong}
\affiliation{Deutsches Elektronen-Synchrotron, 15738 Zeuthen, Germany}%

\author{Xiangkun Li}
\affiliation{Deutsches Elektronen-Synchrotron, 15738 Zeuthen, Germany}%

\author{Osip Lishilin}
\affiliation{Deutsches Elektronen-Synchrotron, 15738 Zeuthen, Germany}%

\author{Zakaria Aboulbanine}
\affiliation{Deutsches Elektronen-Synchrotron, 15738 Zeuthen, Germany}%

\author{Gowri Adhikari}
\affiliation{Deutsches Elektronen-Synchrotron, 15738 Zeuthen, Germany}%

\author{Namra Aftab}
\affiliation{Deutsches Elektronen-Synchrotron, 15738 Zeuthen, Germany}%

\author{Prach Boonpornprasert}
\affiliation{Deutsches Elektronen-Synchrotron, 15738 Zeuthen, Germany}%

\author{Georgi Georgiev}
\affiliation{Deutsches Elektronen-Synchrotron, 15738 Zeuthen, Germany}%

\author{James Good}
\affiliation{Deutsches Elektronen-Synchrotron, 15738 Zeuthen, Germany}%

\author{Matthias Gross}
\affiliation{Deutsches Elektronen-Synchrotron, 15738 Zeuthen, Germany}%

\author{Christian Koschitzki}
\affiliation{Deutsches Elektronen-Synchrotron, 15738 Zeuthen, Germany}%

\author{Raffael Niemczyk}
\affiliation{Deutsches Elektronen-Synchrotron, 15738 Zeuthen, Germany}%

\author{Anne Oppelt}
\affiliation{Deutsches Elektronen-Synchrotron, 15738 Zeuthen, Germany}%

\author{Guan Shu}
\affiliation{Deutsches Elektronen-Synchrotron, 15738 Zeuthen, Germany}%

\author{Frank Stephan}
\affiliation{Deutsches Elektronen-Synchrotron, 15738 Zeuthen, Germany}%

\author{Grygorii Vashchenko}
\affiliation{Deutsches Elektronen-Synchrotron, 15738 Zeuthen, Germany}%

\author{Tobias Weilbach}
\affiliation{Deutsches Elektronen-Synchrotron, 15738 Zeuthen, Germany}%

\date{\today}

\begin{abstract}
Slice energy spread is one of the key parameters in free electron laser optimizations, but its accurate measurement is not straightforward. Two recent studies from high energy ($>$100 MeV) photoinjectors at SwissFEL and European XFEL have reported much higher slice energy spread than expected at their XFEL working points (200 - 250 pC). In this paper, a new method for measuring slice energy spread at a lower beam energy ($\sim$20 MeV) is proposed and demonstrated at the PhotoInjector Test facility at DESY Zeuthen (PITZ), and the results for 250 pC and 500 pC are much lower than those measured at high energy injectors.
\end{abstract}

\maketitle


\section{Introduction}
High brightness electron beams are critical for many scientific instruments, such as electron microscopes and X-ray free electron lasers (XFEL), which have transformed modern research with unprecedented spatial, temporal and energy resolutions \cite{williams1996transmission, pellegrini2017x, bostedt2016linac, rossbach201910}. Beam brightness scales as beam density in the 6 dimensional (6D) phase space
\begin{eqnarray}\label{eq:B6D}
  B_{6D} \propto  \frac{\partial^3 Q}{\partial \varepsilon_{nx} \partial \varepsilon_{ny} \partial \varepsilon_{nz}} \propto \dfrac{I/\sigma_{\scriptscriptstyle E}}{\varepsilon_{nx}\varepsilon_{ny}},
\end{eqnarray}
where $Q$ is the bunch charge, $I$ is the peak current, $\sigma_{\scriptscriptstyle E}$ is the uncorrelated energy spread or slice energy spread, $\varepsilon_{nx}$, $\varepsilon_{ny}$ and $\varepsilon_{nz}$ are normalized emittances. While the beam brightness directly measures the performance of electron microscopes, it indirectly determines the  XFEL brilliance by affecting the amplification gain in the undulator \cite{huang2007review}. Therefore, measurement of 6D beam brightness is an indispensable part of XFEL optimizations. For linac based XFELs, beam brightness optimization has to start from the injector due to the Liouville's theorem. Both transverse emittance and beam peak current can be routinely measured in an XFEL injector, but accurate slice energy spread measurement is still not trivial. This is because photoinjector slice energy spread is expected to be on the 1 keV level, which is below the measurement resolution \cite{huning2003measurement}. In the past, accurate slice energy spread measurement for XFEL injector was not that critical, because it is much lower than required and even causes micro bunching instability to 
reduce XFEL lasing \cite{saldin2002klystron, huang2002formulas, ratner2015time, qiang2017start}. Therefore, a laser heater is used to increase the injector slice energy spread to damp such an instability to improve FEL lasing \cite{huang2010measurements}. With the improvement of electron source brightness and undulator technology, compact XFEL machines of lower beam energy were built and proposed, such as SACLA and SwissFEL \cite{pile2011first, prat2020compact, rosenzweig2020ultra, prat2020compact}. Besides, the injector peak current is also getting lower to improve transverse emittance \cite{filippetto2014maximum}. Due to the lower injector peak current and lower undulator beam energy, the threshold for injector slice energy spread gets much lower. 

Recently, two dispersion based methods were proposed and demonstrated for accurate injector slice energy spread measurements with rms uncertainty down to the 0.1 to 0.3 keV level \cite{PSI:dE, XFEL:dE}. Both measured much higher slice energy spread than expected for 200 to 250~pC bunch charge, which reduced the laser heater role in improving XFEL lasing \cite{bettoni2020impact}. Compared to SASE XFEL, seeded XFEL prefers even lower slice energy spread for high harmonic generation \cite{yu2003first, stupakov2009using}, in which slice energy spread and energy modulation are typically measured with undulator radiaiton dependence over chicane strength after the modulator \cite{feng2011measurement, penco2015experimental}. A summary of high resolution slice energy spread measurements is shown in Table \ref{tab:dE_summary}, and the two hard X-ray injectors show much lower longitudinal beam brightness than the two seeded FEL injectors.

\begin{table}[!htp]
\caption{\label{tab:dE_summary}Summary of slice energy spread measurements.}
\begin{ruledtabular}
\begin{tabular}{cccccc}
  & SDUV & FERMI & SwissFEL & Eu-XFEL & Unit \\\hline
         Q  & 100 & 600 & 200 & 250 & pC \\
$\rm{E_k}$  & 136 & 1320 & 100 & 130 & MeV \\
       $I$  & 12 & 800 & 20 & 20 & A \\
$\sigma_{\scriptscriptstyle E}$  & 1.2 & 40 & 15 & 5.9 & keV \\
$I/\sigma_{\scriptscriptstyle E}$ & 10 & 20 & 1.3 & 3.4 & A/keV \\
Method & \multicolumn{2}{c}{Undulator radiation} & \multicolumn{2}{c}{Dispersion} & /\\
Reference & \cite{feng2011measurement} & \cite{penco2020enhanced} & \cite{PSI:dE} & \cite{XFEL:dE} & /\\
\end{tabular}
\end{ruledtabular}
\end{table}

In this paper, we introduce a dispersion based method to measure the slice energy spread at the low energy (20 MeV) PhotoInjector Test facility at DESY Zeuthen (PITZ). In addition to the low beam energy, tungsten slit masks are used to reduce the beam transverse emittance, therefore energy spread measurement resolution improves compared to a high energy injector. The paper is organized as follows. First, the methodology is described in Sec. \ref{method}. Then, the experiments for 250 pC and 500 pC are presented in Sec. \ref{measurements}. Finally, a discussion and summary is given in Sec. \ref{discussions} and \ref{summary}.

\section{Methodology}
\label{method}
Slice energy spread is usually measured with an RF transverse deflecting structure (TDS) and a dipole magnet, which maps the longitudinal phase space (LPS) of the beam to the transverse distribution on a screen in a dispersion section. As was discussed in \cite{PSI:dE}, the beam size along the energy dispersion direction consists of four contributions, i.e. screen spatial resolution, transverse emittance effect, TDS-induced energy spread and true slice energy spread. The convolution of the four contributions can be expressed as
\begin{eqnarray}\label{eq:total_dE}
  \sigma_{\rm total}^2 = \sigma_{\rm scr}^2 + \dfrac{\varepsilon_{\rm n1}\beta_{\rm scr}}{\gamma} + \left(D\dfrac{\sigma_{\gamma}}{\gamma}\right)^2 + \left(D\dfrac{\sigma_{\gamma,\scriptscriptstyle \rm TDS}}{\gamma}\right)^2,
\end{eqnarray}
where $\sigma_{\rm total}$ is the total rms beam size, $\sigma_{\rm scr}$ is the rms screen resolution, $\varepsilon_{\rm n1}$ is the normalized slice emittance in the dipole bending plane, $\beta_{\rm scr}$ is the beta function at measurement screen, $D$ is the dispersion function, $\gamma$ is the beam Lorentz factor, $\sigma_{\gamma,\scriptscriptstyle \rm TDS}$ is the slice energy spread due to the transverse acceleration gradient in TDS, $\sigma_{\gamma}$ is the net slice energy spread without the contribution of TDS.

The energy spread $\sigma_{\gamma,\scriptscriptstyle \rm TDS}$ depends linearly on the TDS deflection voltage, therefore it can be extracted by TDS voltage scan. After removing the TDS contribution, there are two methods to extract the net slice energy spread $\sigma_{\gamma}$ from the other contributions. The first is demonstrated at the SwissFEL by scanning the beam energy $\gamma$ \cite{PSI:dE}. The second is demonstrated at the European XFEL by scanning the dispersion function $D$ \cite{XFEL:dE}. Both methods require a constant slice emittance and constant beta function at the measurement screen, which is not easy to achieve for a space charge dominated low energy photoinjector like PITZ.

\subsection{Energy spread resolutions: low energy vs high energy}

The photoinjector slice energy spread is very low in free electron laser applications, expected to be few keV from simulations \cite{slac:heater}. To reduce the measurement error of $\sigma_{\gamma}$, the contributions from the other three terms in Eq.~(\ref{eq:total_dE}) cannot be much larger than $\sigma_{\gamma}$, otherwise it puts a tight requirement on machine stability and other measurement errors. Therefore, contributions from screen resolution, transverse emittance and TDS should be as low as possible. The energy spread resolution due to screen resolution and transverse emittance can be expressed as
\begin{eqnarray}\label{eq:scr_dE}
  \sigma_{\gamma,\rm scr} = \dfrac{\sigma_{\rm scr}}{D}\gamma,
\end{eqnarray}
\begin{eqnarray}\label{eq:emit_dE}
  \sigma_{\gamma,\varepsilon} = \dfrac{\sqrt{\varepsilon_{\rm n1}\beta_{\rm scr}}}{D}\gamma^{\frac{1}{2}}.
\end{eqnarray}
The best energy and time resolutions by TDS in the linear approximation are \cite{TDS:dtdE}
\begin{eqnarray}\label{eq:TDS_dE}
  \sigma_{\gamma,\scriptscriptstyle \rm TDS} = \dfrac{e}{m_0c^3}\sqrt{\varepsilon_{\rm n2}\beta_{\scriptscriptstyle \rm TDS}}\dfrac{\omega_{\scriptscriptstyle \rm TDS}V_{\scriptscriptstyle \rm TDS}}{\gamma^{\frac{1}{2}}},
\end{eqnarray}
\begin{eqnarray}\label{eq:TDS_dt}
  \sigma_{t} = \dfrac{m_0c^2}{e}\dfrac{\sqrt{\varepsilon_{\rm n2}/\beta_{\scriptscriptstyle \rm TDS}}}{\sin(\psi)}\dfrac{\gamma^{\frac{1}{2}}}{\omega_{\scriptscriptstyle \rm TDS}V_{\scriptscriptstyle \rm TDS}},
\end{eqnarray}
\begin{eqnarray}\label{eq:TDS_dtdE}
  \sigma_{t}\sigma_{\gamma,\scriptscriptstyle \rm TDS} = \dfrac{\varepsilon_{\rm n2}}{c\sin(\psi)},
\end{eqnarray}
where $e$ is elementary charge, $c$ is speed of light, $\varepsilon_{\rm n2}$ is the normalized slice emittance in the TDS streaking plane, $\beta_{\scriptscriptstyle \rm TDS}$ is the beam beta function in the TDS, $\omega_{\scriptscriptstyle \rm TDS}$ is the TDS angular frequency, $V_{\scriptscriptstyle \rm TDS}$ is the TDS transverse voltage, $\psi$ is the phase advance between TDS and the LPS measurement screen.

In case of same $\sigma_{\rm scr}$, $D$, $\varepsilon_{\rm n}$, $\beta_{\rm scr}$, Eq.~(\ref{eq:scr_dE}) and Eq.~(\ref{eq:emit_dE}) show better absolute energy spread resolution can be achieved with a lower beam energy. For LPS measurements, its best time and energy resolutions product is limited by the transverse emittance and phase advance only. A better time resolution from TDS will lead to a worse energy spread resolution for the LPS measurement.

Taking the parameters from the European XFEL injector as an example \cite{XFEL:dE}, beam energy is 130 MeV, dipole screen resolution is 28 \textmu m with an energy dispersion of 1.2 m, nominal emittance is 0.4 \textmu m for 250 pC \cite{xfel250pC2020, XFEL:dE}. The screen-induced energy spread resolution is 3 keV, but it can be reduce to 0.5 keV if the beam energy is lowered to 20 MeV. The product of TDS time resolution and energy resolution is at least 0.68 keV ps based on Eq.~(\ref{eq:TDS_dtdE}). If temporal resolution of TDS measurement reaches a 1 ps (FWHM) resolution, i.e. 0.42 ps rms, TDS will induce an rms energy spread of at least 1.6 keV.

\subsection{TDS voltage scan}
Since $\sigma_{\gamma,\scriptscriptstyle TDS}$ is linearly proportional to $V_{\scriptscriptstyle TDS}$, Eq.~(\ref{eq:total_dE}) can be rewritten as
\begin{eqnarray}\label{eq:total_dE2}
  \sigma_{\rm total}^2 = \sigma_0^2 + (\sigma_1\dfrac{V_{\scriptscriptstyle \rm TDS}}{V_1})^2,
\end{eqnarray}
where $\sigma_0^2$ is the sum of first three terms on the right hand side of Eq.~(\ref{eq:total_dE}), which are independent of TDS voltage. $\sigma_1$ is the beam size contribution from TDS voltage $V_1$. With a scan of $\sigma_{\rm total}$ versus $V_{\scriptscriptstyle \rm TDS}$ in experiment, $\sigma_0$ can be fitted.

\begin{figure}[!thp]
   \includegraphics*[width=\columnwidth]{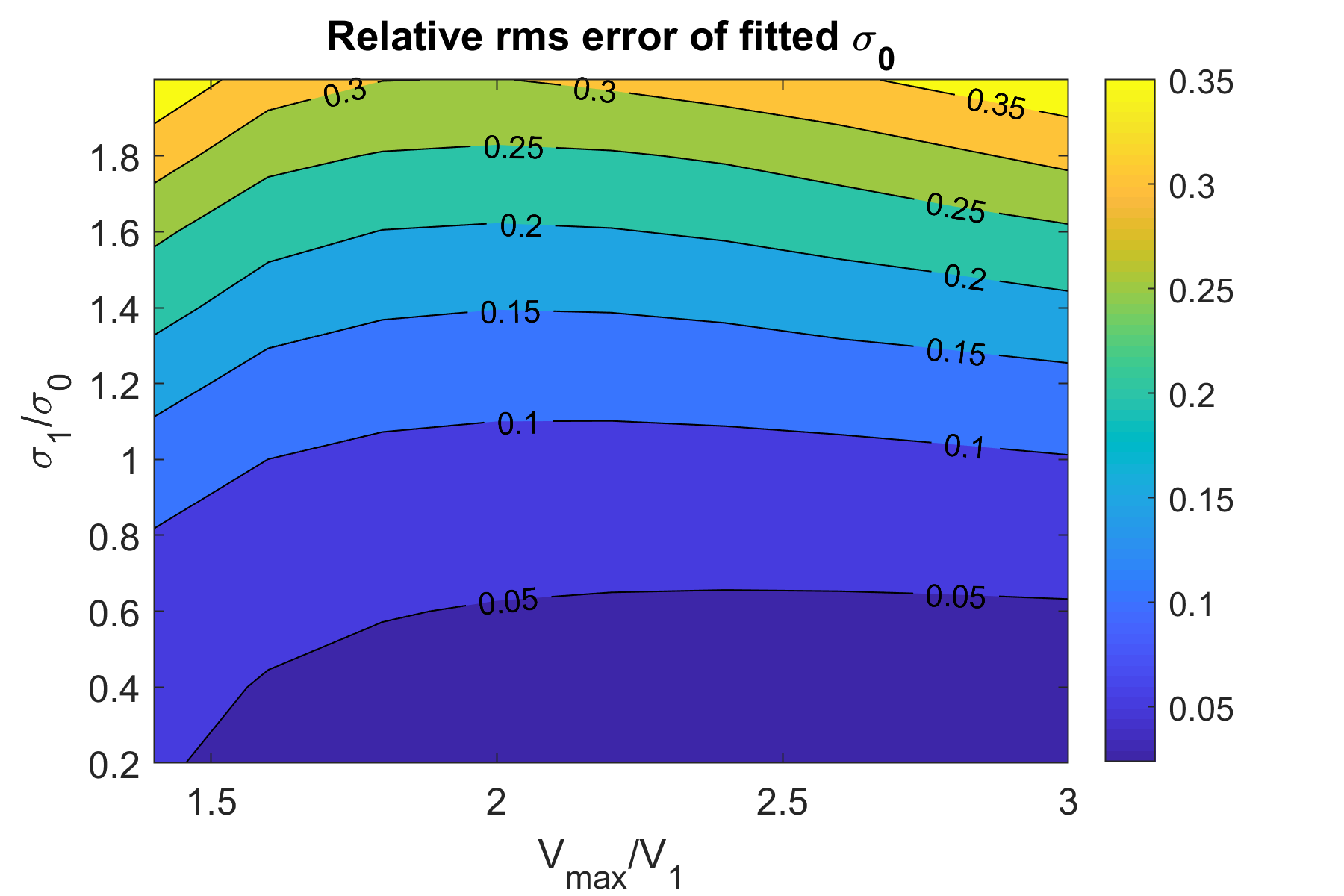}
   \caption{\label{fig:fit_error}Relative rms error of $\sigma_0$ vs TDS voltage scan configuration.}
\end{figure}

Let us assume a 3\% rms error for $\sigma_{\rm total}$ measurement.
Here an numerical example is used to show the sensitivity of $\sigma_0$ fitting error on TDS voltage scan range. Let us assume TDS voltage scan has 6 voltages uniformly distributed between $V_1$ and $V_{\rm max}$. For each TDS voltage, $\sigma_{\rm total}$ is calculated based on Eq.~(\ref{eq:total_dE2}), and a random relative error with a 3\% rms value is added. With 6 TDS voltages and corresponding energy spread values, $\sigma_0$ can be fitted. Such a process is repeated 1000 times, and the rms relative errors are shown in Fig.~\ref{fig:fit_error} for different measurement configurations. The simulations show that the fitting error depends critically on $\sigma_1$, i.e. $V_1$. To keep the fitting error of $\sigma_0$ below 10\%, $\sigma_1$ should be smaller than $\sigma_0$. With the TDS voltage scan range $V_{\rm max}/V_1$ between 1.5 and 3, the fitting error is not sensitive to $V_{max}$. Based on Eq.~(\ref{eq:TDS_dtdE}), a smaller $\sigma_1$ means a worse time resolution, which might increase the measurement error of $\sigma_1$ by including time correlated energy spread. Therefore, a low transverse emittance in the TDS streaking plane is also critical.
\subsection{Slice energy spread measurement in a low energy beamline}
\label{method:lowEk}
Once the slice energy spread from TDS is precisely removed, slice energy spread $\sigma_{\gamma}$ measurements are limited by screen resolution and transverse emittance effect. Energy scan or dispersion scan methods demonstrated for high energy injector at SwissFEL and European XFEL \cite{PSI:dE, XFEL:dE} can't be easily applied to low energy beamlines, where constant slice emittance and beta function are difficult to maintain under the space charge effect. Instead, we propose to measure the screen resolution and emittance-induced beam size directly.

\begin{figure}[!thp]
   \includegraphics*[width=\columnwidth]{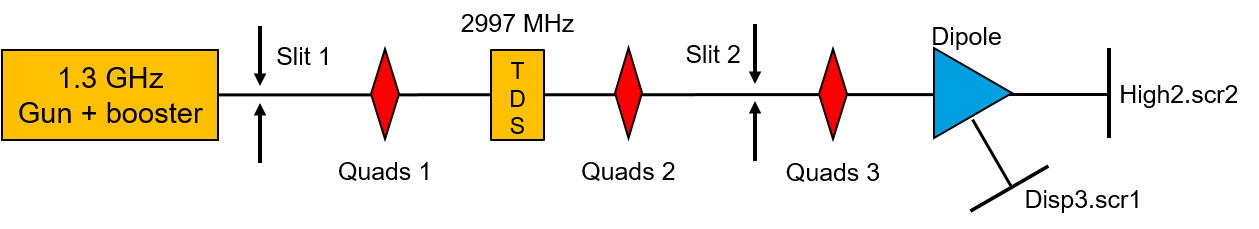}
   \caption{\label{fig:beamline}Schematic beamline for slice energy spread measurement at PITZ (element positions are not in proportion).}
\end{figure}
Figure~\ref{fig:beamline} shows the schematic beamline for slice energy spread measurements at PITZ. A 20 MeV high brightness beam is produced by a 1.3 GHz photoinjector, and sent to a LPS diagnostic section consisting mainly of an S-band TDS and a dipole magnet \cite{TDS:thesis, PITZ:LPS}. The TDS streaks the beam temporally in the vertical direction \cite{PITZ:TDS}, and the dipole bends the beam horizontally. The energy dispersion function is 0.9 m at the LPS measurement screen (Disp3.scr1) \cite{TDS:thesis}. The 1.3 GHz RF gun is the same type gun as the ones used at FLASH and European XFEL \cite{BD1997, FS2010, MK2012}. The photoinjector can produce long bunch trains with micro repetition rate of 1 MHz, but the TDS can only streak up to 3 pulses in a train due to its short RF pulse length. 

There are two slit stations shown in Fig.~\ref{fig:beamline}, and both of them have horizontal slits and vertical slits to cut the beam \cite{staykov2005design}. In this paper, horizontal slit reduces beam vertical emittance, and vertical slit reduces beam horizontal emittance. Slit 1 with a 50 \textmu m opening is used to cut the beam vertical emittance to improve the TDS time and energy resolution, as indicated by Eq.~(\ref{eq:TDS_dtdE}). Quadrupole magnet group 1 tunes the beam size inside the TDS to reduce $\sigma_{\gamma,\scriptscriptstyle \rm TDS}$. Slit station 2 will be used to measure screen resolution and transverse emittance-induced beam size at the LPS measurement screen station Disp3.scr1.

For the Disp3.scr1 resolution measurement, a horizontal slit 2 with 10 \textmu m opening cuts the beam. Since the beam size before slit 2 is much larger than 10 \textmu m, the vertical rms size after the slit is 2.9 \textmu m. Due to such an negligible vertical beam size and negligible space charge effect after the slits, the vertical rms beam size on Disp3.scr1 equals
\begin{eqnarray}\label{eq:scr_reso}
  \sigma_{y}^2 = \sigma_{\rm scr}^2 + (R_{12y}\cdot\sigma_{y', \rm slit2})^2,
\end{eqnarray}
where $R_{12y}$ is the transfer matrix element from slit 2 to Disp3.scr1, $\sigma_{y', \rm slit2}$ is the beam divergence after slit 2 cut. The $R_{12y}$ can be either calculated if a reliable lattice model is established, or measured directly by orbit response. By scanning $R_{12y}$, using quadrupole group 3, and measuring vertical rms size on Disp3.scr1, the screen resolution can be fitted based Eq.~(\ref{eq:scr_reso}).

For measuring emittance-induced horizontal beam size on Disp3.scr1, the vertical slit 2 is used, and the horizontal slit 2 is removed. The final slice energy spread is measured with a combination of horizontal slit 1 and  vertical slit 2. To allow enough charge for the slice energy spread measurement, the vertical slit 2 opening is 50 \textmu m. After slit 2 cut, the beam horizontal rms size is 14.4 \textmu m, which is much smaller than horizontal beam size on High2.scr2 when dipole magnet is off, and can be neglected. So, the horizontal beam size at High2.scr2 equals
\begin{eqnarray}\label{eq:beam_x'}
  \sigma_{x}^2 = \sigma_{\rm scr}^2 + (R_{12x,\rm H2}\cdot\sigma_{x',\rm slit2})^2,
\end{eqnarray}
where $R_{12x,\rm H2}$ is the transfer matrix element from slit 2 to High2.scr2, $\sigma_{x',\rm slit2}$ is the beam divergence after slit 2 cut. Similar to screen resolution measurement, by varying $R_{12x,\rm H2}$, both screen resolution and beam divergence can be fitted. Then beam emittance-induced beam size at Disp3.scr1 can be calculated as $R_{12x,\rm D3}\cdot\sigma_{x',\rm slit2}$, where $R_{12x,\rm D3}$ is the transfer matrix element from slit 2 to Disp3.scr1.

After slice energy spread contributions from TDS, screen resolution and transverse emittance are measured, the real slice energy spread can be extracted via Eq.~(\ref{eq:total_dE}).

\section{Measurements}
\label{measurements}
As mentioned already, recent slice energy spread measurements at the photoinjectors of SwissFEL and European XFEL have shown slice energy spread much higher than expected from simulations \cite{PSI:dE, XFEL:dE}. This is either due to high slice energy spread already from the low energy section, or due to slice energy spread growth during the high energy acceleration and transportation. This motivated us to measure the slice energy spread in the low energy (20 MeV) injector at PITZ, which has the same RF gun as the European XFEL.

In the experiment, the RF gun and the cathode laser mimic the 250 pC working point of the European XFEL injector \cite{chen2020beam, xfel250pC2020, gross:ipac2021-wepab040}. The photoelectron beam is generated by UV laser illuminating the $\rm{Cs_2Te}$ cathode. Cathode laser diameter is 1 mm with a quasi-uniform distribution. Temporally, the laser is 7 ps (FWHM) with a Gaussian distribution. The RF gun accelerates the beam to 6.3 MeV/c with a cathode gradient of 58 MV/m. Then, the beam is matched into the booster linac by solenoid focusing for optimum emittance at the booster exit. Finally, the 20 MeV beam is sent to the diagnostic beamline for slice energy spread measurement. The beam peak current is measured by the TDS to be 20 A.

As discussed in Sec. \ref{method:lowEk}, the high resolution LPS measurement is done with both slit 1 and slit 2. Slit 1 is used to reduce vertical emittance to increase TDS resolution, and slit 2 is used to reduce horizontal emittance-induced energy spread resolution on Disp3.scr1. Both slits have 50 \textmu m opening. In order to measure the low charge beams, 500 µm thick LYSO:Ce is used as screen material at High2.scr2 and Disp3.scr1 \cite{wiebers2013scintillating}. Quadrupole group 1 reduces the vertical beam size in the TDS to lower the TDS-induced energy spread, and quadrupole group 2 optimizes the time resolution of LPS measurement. Quadrupole group 3 varies vertical $R_{12y}$ and horizontal $R_{12x}$ for measuring screen resolution and emittance-induced energy resolution, respectively.

\begin{figure}[!thp]
   \includegraphics*[width=0.8\columnwidth]{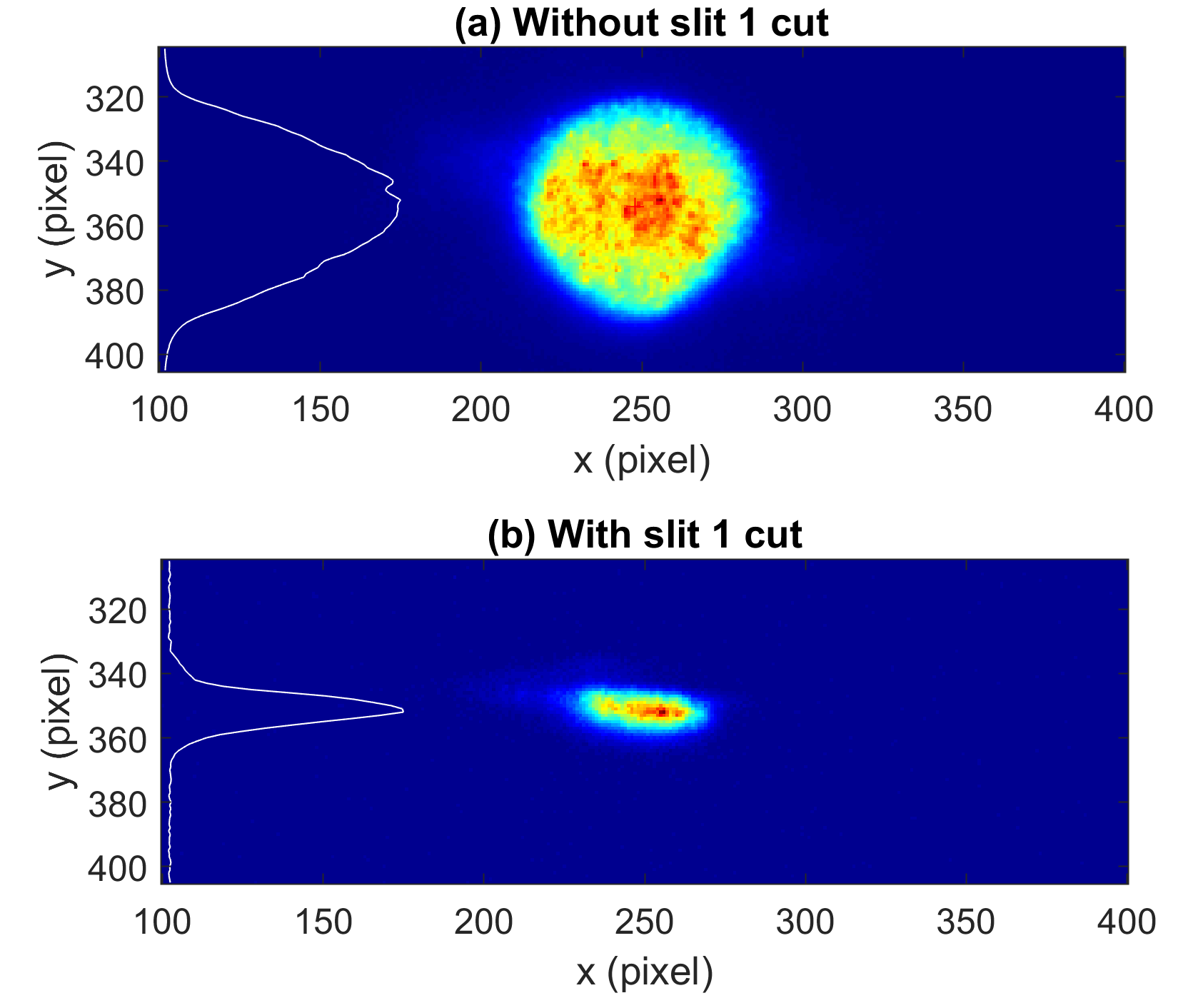}
   \caption{\label{fig:MOIcut}Beam images measured at 3.1 m downstream slit 1, near quadrupole group 1, when quadrupole group 1 is off, (a) without slit 1 cut, (b) with slit 1 cut.}
\end{figure}

The slit 1 effect on vertical beam emittance is shown in Fig.~\ref{fig:MOIcut}. In the optimum emittance compensation working point, the beam is diverging at slit 2, and the normalized vertical rms emittance is around 0.5 \textmu m. The beam profile without slit 1 cut is measured near quadrupole group 1 when they are off, shown in Fig.~\ref{fig:MOIcut} (a). After the slit 1 insertion, the beam profile is shown in Fig.~\ref{fig:MOIcut} (b). The charge of the reduced beamlet is estimated to be 25 pC, and vertical rms size is reduced to 0.17 \textmu m. The drift distance from the slit 1 to the beam profile measurement screen is 3.1 m, so the reduced vertical emittance is calculated to be 32 nm. The vertical emittance is reduced by a factor of 15.6 compared to the nominal case, which improved the LPS resolution significantly. 

\begin{figure}[!thp]
   \includegraphics*[width=0.8\columnwidth]{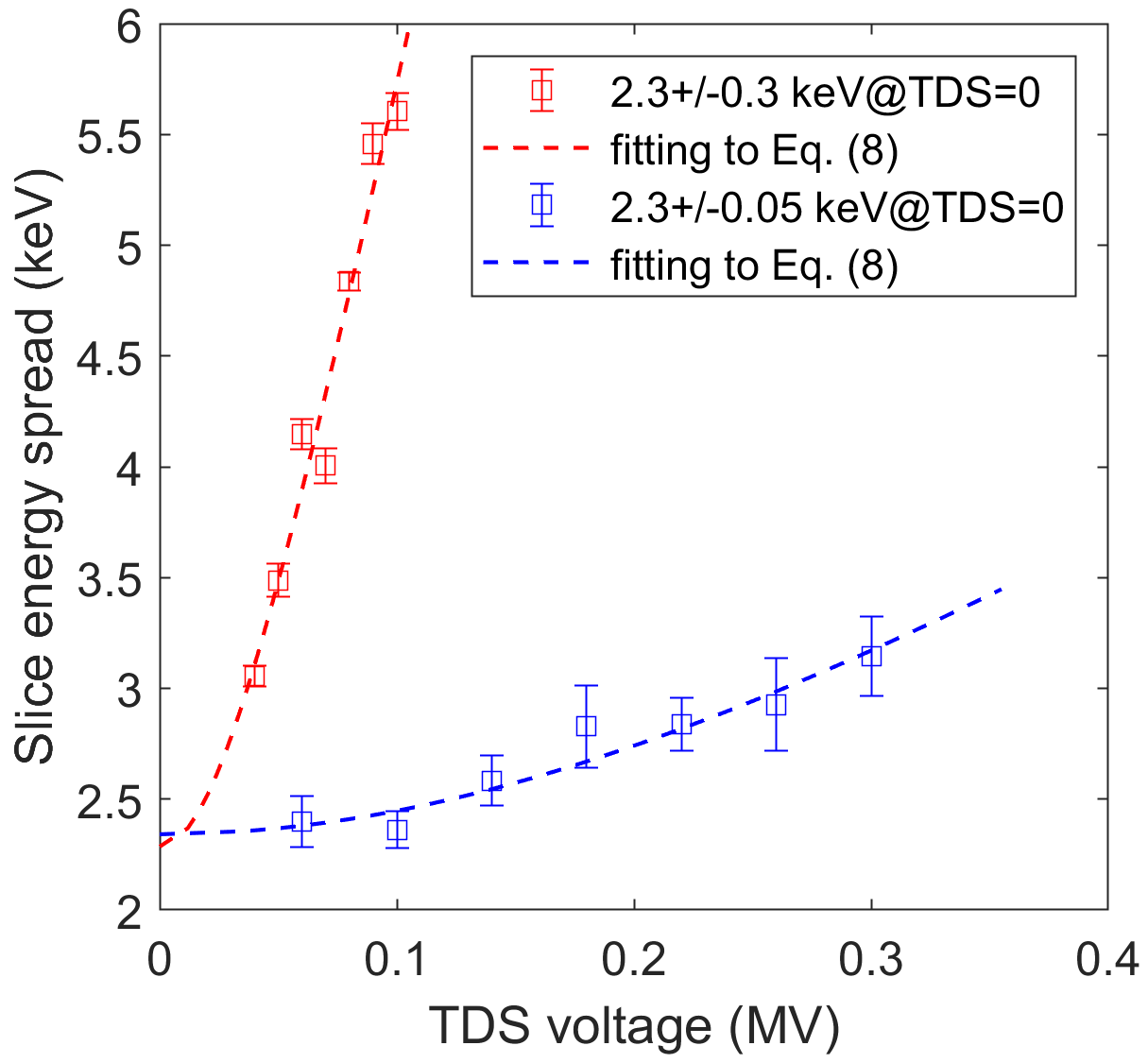}
   \caption{\label{fig:TDS_scan}TDS voltage scan with and without vertical emittance reduction by slit1.}
\end{figure}

\begin{figure}[!thp]
   \includegraphics*[width=0.8\columnwidth]{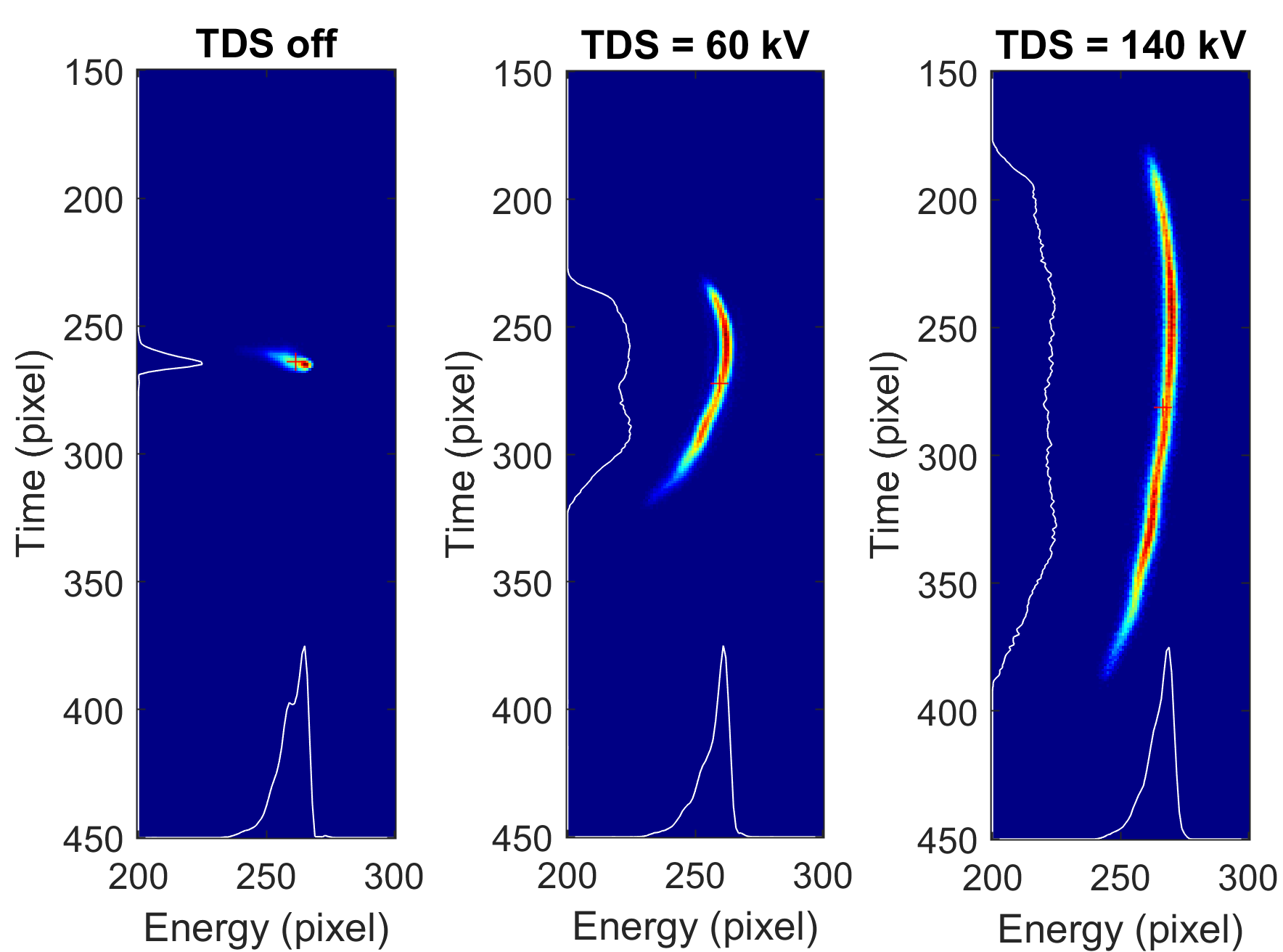}
   \caption{\label{fig:LPS}LPS vs TDS voltage with both slit 1 and slit 2 inserted.}
\end{figure}

The TDS voltage scan with and without slit 1 insertion is shown in Fig.~\ref{fig:TDS_scan}, and the corresponding LPS images are shown in Fig.~\ref{fig:LPS}. The slice energy spread is calculated for the slice at the TDS zero crossing phase. To reduce the correlated energy spread in the slice, the booster linac phase is tuned to minimize linear energy chirp at TDS zero crossing phase. This is very important for the first TDS voltage, whose time resolution is the worse among all TDS voltages. Figure~\ref{fig:TDS_scan} shows the advantage of small vertical emittance after slit 1 cut. Without slit 1 cut, it's difficult to find both good time resolution and low energy spread. The high energy spread at the first TDS voltage increases the fitting error as shown in Fig.~\ref{fig:fit_error}. With the slit 1 insertion, a much smaller vertical emittance reduces the product of TDS time and energy resolution significantly. The vertical beam size in the TDS is reduced by quadrupole group 1 to reduce energy spread growth while a good time resolution can still be maintained at Disp3.scr1. The slice energy spread growth by the first TDS voltage is almost negligible, therefore the rms error is reduced from 0.3 keV to 0.05 keV. Since the final slice energy spread is expected to be on the 1 keV level, such an error reduction is very important. It is worth to mention, 2.3 $\pm$ 0.05 keV still includes contributions from screen resolution and emittance contribution, but it is already much lower than high energy injector results, such as 5.9 $\pm$ 0.1 keV for European XFEL injector at 250 pC and 15 $\pm$ 0.3 keV for SwissFEL injector at 200~pC \cite{PSI:dE, XFEL:dE}. This indicates there might be a slice energy spread growth in the high energy photoinjector.

\begin{figure}[!thp]
   \includegraphics*[width=0.8\columnwidth]{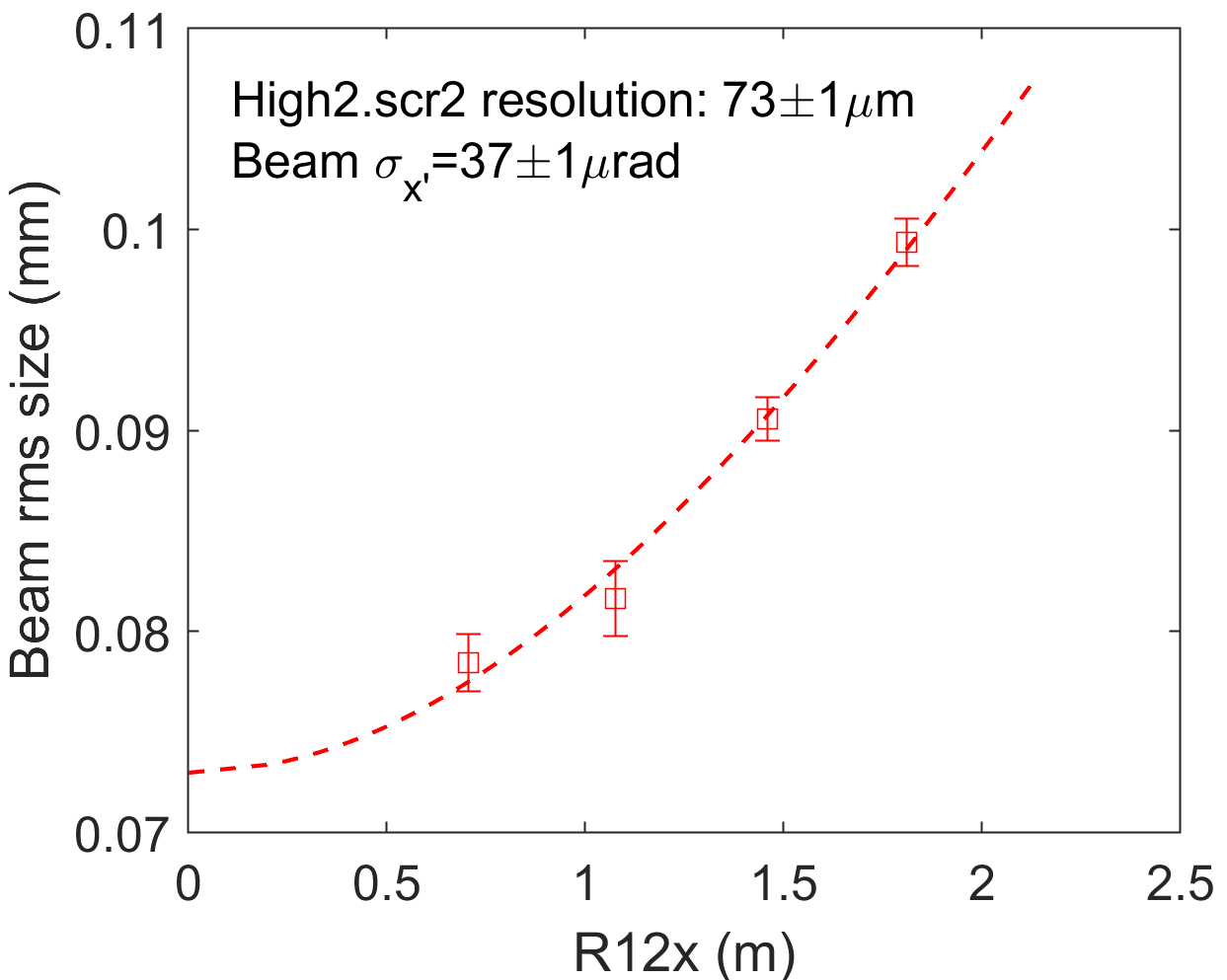}
   \caption{\label{fig:high2scr2}Screen resolution and beam divergence measurements at High2.scr2.}
\end{figure}

As discussed in Sec.~\ref{method:lowEk}, when 50 \textmu m slit 2 vertical slit is inserted, the emittance-induced horizontal rms size on Disp3.scr1 equals $R_{12x,\rm D3}\cdot\sigma_{x',\rm slit2}$. The beam divergence after slit 2 is measured at High2.scr2 when the dipole magnet is off. In order to remove the screen resolution contribution, quadrupole group 3 is used to vary $R_{12x,\rm H2}$, and both the screen resolution and beam divergence can be fitted based on Eq.~(\ref{eq:beam_x'}). Here, the beam rms size $\sigma_x$ at High2.scr2 is measured for the same time slice as that used for slice energy spread calculation. The measurement results are displayed in Fig.~\ref{fig:high2scr2}. When quadrupole group 3 is off, the measured beam size on High2.scr2 is 99~$\pm$~1~\textmu m, but this is dominated by the screen resolution of 73~$\pm$~1~\textmu m. Therefore, the real beam size is only 67~$\pm$~1~\textmu m, which translates to a beam divergence of 37~$\pm$~1~\textmu rad, corresponding to a horizontal normalized emittance of 21~nm. $R_{12x,\rm D3}$ is measured by orbital response to be 0.81~m, so the emittance-induced energy spread resolution is 0.65~$\pm$~0.02~keV.
\begin{figure}[!thp]
   \includegraphics*[width=0.8\columnwidth]{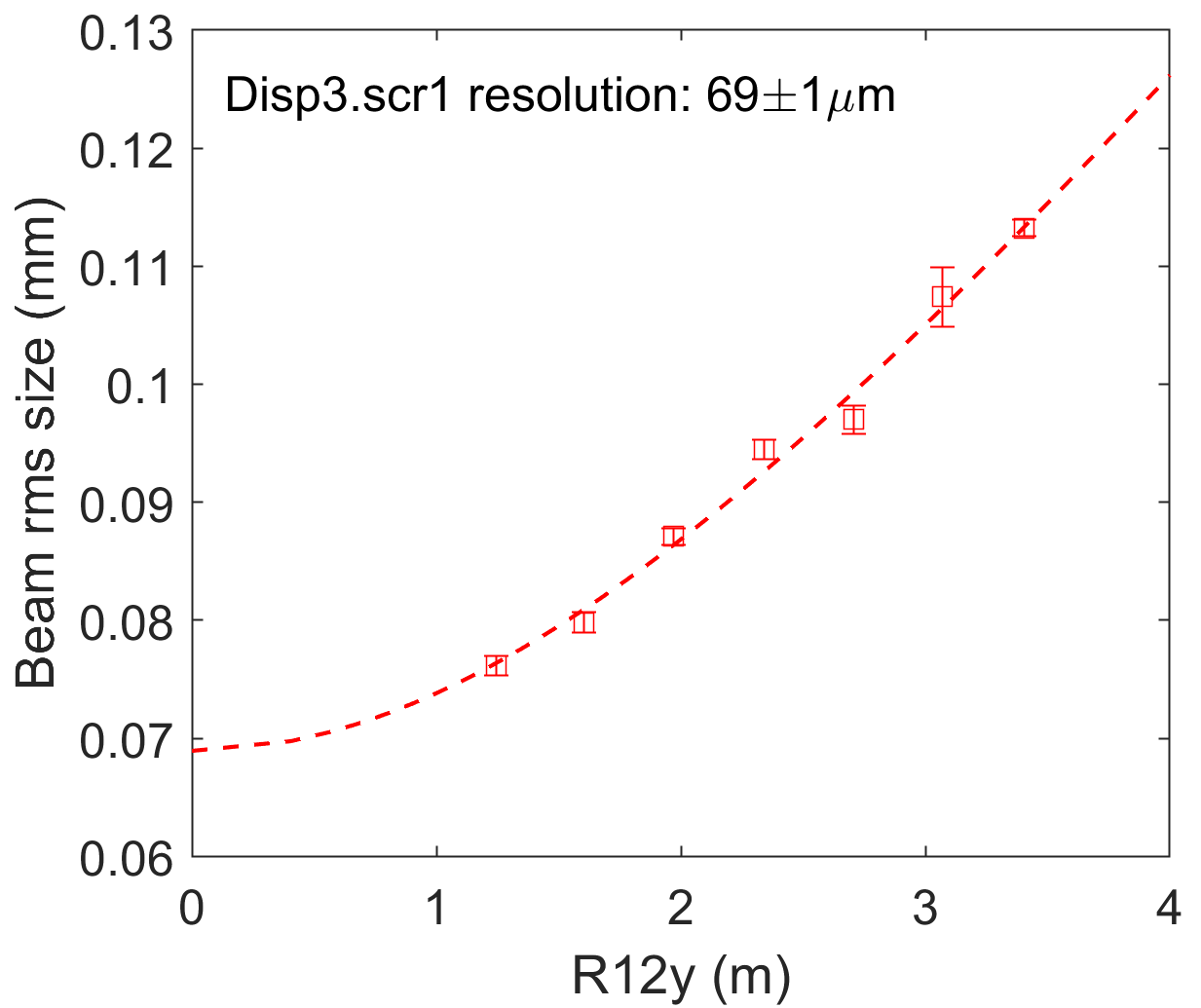}
   \caption{\label{fig:Disp3scr1}Screen resolution measurements at Disp3.scr1.}
\end{figure}
The screen resolution of Disp3.scr1 is measured at the non-dispersion direction when TDS is off, so vertical slit 2 is changed to horizontal slit 2. The measurement is based on Eq.~(\ref{eq:scr_reso}), and $R_{12y}$ is varied by quadrupole group 3. The screen resolution is 69 $\pm$ 1 \textmu m as shown in Fig.~\ref{fig:Disp3scr1}, and this translates to an energy spread resolution of 1.50 $\pm$ 0.02~keV.

With both screen resolution and emittance contribution measured, the slice energy spread deconvolutions are summarized in Table \ref{tab:dE250}. The slice energy spread for the 250 pC beam is 1.65 $\pm$ 0.06 keV. The longitudinal peak brightness, defined by the ratio of peak current over slice energy spread, equals 12.1 A/keV.
\begin{table}[!thp]
\caption{\label{tab:dE250}Summary of 250 pC slice energy spread measurement components, both in beam size measured at Disp3.scr1 and in energy spread.}
\begin{ruledtabular}
\begin{tabular}{ccccc}
 Non-TDS & Screen & Emittance & Real slice & Unit \\
 contribution & resolution & resolution & energy spread\\ \hline
 107 $\pm$ 2 & 69 $\pm$ 1 & 30 $\pm$ 1 & 76 $\pm$ 2 &\textmu m \\
 2.32 $\pm$ 0.05 & 1.50 $\pm$ 0.02 & 0.65 $\pm$ 0.02 & 1.65 $\pm$ 0.06 &keV \\
\end{tabular}
\end{ruledtabular}
\end{table}

Slice energy spread of the 500 pC beam was also measured with the same methods. The cathode laser diameter was changed to 1.3 mm for the best transverse emittance. The other configurations, such as laser temporal profile, gun and booster energy, were the same as for the 250 pC beam. The best emittance is about 0.65 \textmu m with a peak current of 34 A. The slice energy spread deconvolutions are summarized in Table \ref{tab:dE500}. The slice energy spread for the 500 pC beam is 2.84 $\pm$ 0.07 keV, and the longitudinal peak brightness equals 12.0 A/keV.
\begin{table}[!thp]
\caption{\label{tab:dE500}Summary of 500 pC slice energy spread measurement components, both in beam size measured at Disp3.scr1 and in energy spread.}
\begin{ruledtabular}
\begin{tabular}{ccccc}
 Non-TDS & Screen & Emittance & Real slice & Unit \\
 contribution & resolution & resolution & energy spread\\ \hline
 149 $\pm$ 3 & 69 $\pm$ 1 & 15 $\pm$ 1 & 131 $\pm$ 3 &\textmu m \\
 3.23 $\pm$ 0.06 & 1.50 $\pm$ 0.02 & 0.32 $\pm$ 0.03 & 2.84 $\pm$ 0.07 &keV \\
\end{tabular}
\end{ruledtabular}
\end{table}

\section{Discussions}
\label{discussions}
For the 250 pC working point, 1.65 $\pm$ 0.06 keV slice energy spread was measured at the low energy ($\sim$20 MeV) photoinjector at PITZ. This is a factor of 3.6 lower than that measured at high energy (130 MeV) photoinjector at European XFEL, and a factor of 9 lower than that measured at SwissFEL injector for the 200 pC working point. All three injectors operate with the $\rm{Cs_2Te}$ photocathode, and have similar peak current of 20 A. Our result demonstrates the expected 1-2 keV energy spread from the $\rm{Cs_2Te}$ based photoinjector for the first time. This indicates a slice energy spread growth in the high energy photoinjector, which is worth further studies \cite{XFEL:dE}.

ASTRA simulations of the PITZ photoinjector shows 1.3 keV and 2 keV slice energy spread for the 250 pC and the 500 pC respectively \cite{astracode}. Simulation values are lower than measurement results by 20\% to 30\%. This discrepancy can be caused by laser temporal modulations \cite{musumeci2011nonlinear}, or simplified physics model in ASTRA simulations, such as intra beam scattering effect \cite{XFEL:dE}, self consistent space charge modelling close to the saturation photoemission regime \cite{ciocci1997self, chen2018modeling,krasilnikov2019studies}.

\section{Summary}
\label{summary}
Longitudinal phase space mapping by TDS and dipole magnet is used for direct slice energy spread measurement, but its energy resolution is limited by screen resolution, emittance-induced beam size and TDS-induced energy spread. Analytical analysis shows, a low energy beam can achieve better energy resolution than high energy beam with the same normalized emittance, beta function, dispersion function and screen resolution. The product of time and energy resolution of TDS streaking is limited by the beam emittance. Therefore, we used a beam of lower energy and reduced horizontal and vertical emittance by slit cutting to enhance the LPS resolutions. 

Numerical simulations show the lowest TDS-induced energy spread during a TDS voltage scan should be as small as possible to minimize the fitting uncertainty of non-TDS related energy spread. Direct measurement of energy spread resolutions due to screen resolution and emittance-induced beam size were suggested and demonstrated by using slit and lattice scans, which does not require a constant beta function like previous methods used for high energy injectors \cite{PSI:dE, XFEL:dE}.

Finally, we demonstrated the new methods by measuring the slice energy spread at PITZ, which mimic the best emittance working point for European XFEL at 20 MeV. The slice energy spread results are 1.65 $\pm$ 0.06 keV and 2.84 $\pm$ 0.07 keV for 250 pC and 500 pC, respectively. The value for 250 pC is a factor of 3.6 to 9 lower than high energy injector results at European XFEL and SwissFEL, indicating slice energy spread growth in the high energy injector, which is worth further studies. 

\section{Acknowledgements}
The authors appreciate discussions with S. Tomin and I. Zagorodnov, and encouragements from the European XFEL beam dynamics team. This work was supported by the European XFEL research and development program.
\bibliography{apssamp}

\end{document}